\documentclass{article}
\usepackage{spconf,amsmath,graphicx,hyperref}
\usepackage[OT1]{fontenc}
\usepackage{amssymb,bm}
\usepackage{graphicx}
\usepackage{capt-of}
\usepackage{xcolor}
\hypersetup{hidelinks}
\AtBeginDocument{%
  \setlength{\abovedisplayskip}{6pt plus 2pt minus 2pt}%
  \setlength{\belowdisplayskip}{6pt plus 2pt minus 2pt}%
  \setlength{\abovedisplayshortskip}{0pt plus 2pt}%
  \setlength{\belowdisplayshortskip}{3pt plus 2pt minus 1pt}%
}
\makeatletter
\newcommand{\compactfigurecaptions}{%
  \long\def\@makecaption##1##2{%
    \vskip 4pt
    \setbox\@tempboxa\hbox{##1. ##2}%
    \ifdim\wd\@tempboxa>\hsize
      ##1. ##2\par
    \else
      \hbox to\hsize{\hfil\box\@tempboxa\hfil}%
    \fi
  }%
}
\makeatother

\newcommand{\cQ}{\mathcal{Q}}
\newcommand{\cV}{\mathcal{V}}
\newcommand{\CN}{\mathcal{CN}}
\newcommand{\E}{\mathbb{E}}
\DeclareMathOperator{\diag}{diag}

\title{BLACK-BOX ATTACK FOR IRS-AIDED COMMUNICATIONS WITH BER-ONLY FEEDBACK}
\name{%
  Zhengkai Tu$^{1}$,\quad
  Xinyue Xiang$^{2}$,\quad
  Jimmy Huang$^{3,}$\sthanks{\fontsize{9}{10}\selectfont Corresponding author: Jimmy Huang (jhuang@yorku.ca).}%
}
\address{%
  \normalsize
  \begin{tabular}[t]{@{}c@{}}
    $^{1}$The Chinese University of Hong Kong\qquad
    $^{2}$University of Michigan\qquad
    $^{3}$York University
  \end{tabular}%
}
\begin{document}
\raggedbottom
\maketitle
\begin{abstract}
Intelligent reflecting surface (IRS) has emerged as a promising wireless technology for enhancing legitimate communications. In this paper, we investigate an IRS-assisted multiuser MISO downlink network in which an attacker reconfigures the IRS to degrade legitimate data transmission. Specifically, we consider a black-box attack setting in which the attacker has access to neither channel information nor symbol-level observations and can observe only the BER reported by each user. The objective of the attacker is to identify an IRS configuration that maximizes the minimum BER among all users under a limited budget of non-repeated tests. To address this BER-only black-box attack problem, we propose PRIS, an online search method that learns effective phase transitions from previous trials and progressively narrows the search from large-block modifications to single-element refinement. Specifically, an annealed user-balancing reward and prioritized replay are incorporated to exploit noisy BER feedback. Numerical results demonstrate that PRIS achieves a higher minimum BER than existing benchmarks.
\end{abstract}
\begin{keywords}
Intelligent reflecting surface (IRS), BER-only feedback, black-box attack, multiuser MISO
\end{keywords}

\section{Introduction and Motivation}
\label{sec:introduction}

An intelligent reflecting surface (IRS), also known as a reconfigurable intelligent surface (RIS), enhances wireless communication by coordinating reflected signal phases~\cite{wu2021tutorial}. Its programmable response supports a range of communication applications~\cite{gong2020toward}, including energy-efficient transmission through joint phase-shift and power allocation design~\cite{huang2019reconfigurable}. This controllability creates new opportunities for wireless system design~\cite{liu2021reconfigurable}. Communication-theoretic analyses characterize RIS operating principles and performance limits~\cite{basar2019ris}. Physics-based models complement these analyses by relating surface responses to smart radio environments~\cite{direnzo2020smart}. We study IRS attacks using BER-only feedback.

Most existing studies use the IRS to enhance legitimate transmission. Joint active and passive beamforming addresses transmit-power minimization in multiuser MISO systems~\cite{wu2019joint}, weighted sum-rate maximization~\cite{guo2020wsr}, and multicell MIMO communication~\cite{pan2020multicell}. Practical designs incorporate discrete phase-shift constraints~\cite{wu2020discrete}. The IRS can also protect legitimate transmission through reinforcement learning for adaptive power allocation and beamforming~\cite{yang2021antijamming}, or through robust designs that account for incomplete jammer information~\cite{sun2022outage}.

When channel state information (CSI) is unavailable, received-power measurements can guide blind beamforming. In particular, Ren et al.\ select IRS phases using conditional sample means of received power~\cite{ren2023blind}. Dinh-Van et al.\ instead learn a quadratic signal-to-noise ratio model from power measurements to optimize binary phases~\cite{dinhvan2025born}. Blind passive beamforming has also been studied for MIMO systems~\cite{lai2025blindmimo}.

Malicious IRS control can degrade legitimate transmission through destructive signal combining without active transmit power~\cite{lyu2020jamming} or through signal leakage and reflected interference~\cite{wang2022illegal}, motivating secure transmission schemes in the presence of illegal RISs~\cite{shao2023secure}. Other attacks target channel acquisition through pilot contamination~\cite{huang2021pilot} or pilot spoofing based on statistical CSI~\cite{yang2021spoofing}. Recent work optimizes non-diagonal RIS attacks using GA--DRL and channel-distribution information~\cite{li2025gadrl}. Alternatively, CSI-free random reconfiguration of omni-surfaces induces active channel aging~\cite{huang2026discoios}. This mechanism has also been studied with distributed IRSs~\cite{wang2026distributeddisco}. MALRIS addresses the complementary threat of compromised RIS hardware and firmware~\cite{mughal2025malris}.

In contrast to power-based blind beamforming and random reconfiguration attacks, we seek a single fixed discrete IRS configuration using only users' BER feedback. We consider a multiuser MISO downlink whose legitimate IRS configuration and ZF precoder have been optimized. An undetected attacker modifies the IRS while the ZF precoder remains fixed. The attacker has no access to CSI or symbol-level information. Each complete phase vector can be tested only once under a finite query budget. The objective is to maximize the minimum BER across users. To this end, we propose Progressive Reward-Informed Search (PRIS), which learns from previously tested configurations and BER feedback to guide a progressive coarse-to-fine search.

\section{System Model and Problem Formulation}
\label{sec:system_problem}

We consider an IRS-assisted multiuser MISO downlink system, where
an $N_t$-antenna base station (BS) communicates with $U$ single-antenna users with the assistance of an
IRS with $N$ reflective elements (REs). Meanwhile, an unknown attacker attempts to disrupt normal data transmission by manipulating the IRS configuration. We first describe the data transmission model and then introduce the attack model. Let $\theta_n$ be the phase shift induced by the $n$-th RE. From a practical standpoint, each RE can apply one of $K$ phase shifts from the discrete set
\begin{equation}
 \Phi_K\triangleq\left\{\frac{2\pi\ell}{K}\right\}_{\ell=0}^{K-1}.
 \label{eq:phase_space}
\end{equation}
Let $ \bm\theta=[\theta_1,\ldots,\theta_N]$ be the phase shift vector and the reflection
matrix is $\bm\Theta(\bm\theta)=\diag(e^{j\theta_1},\ldots,e^{j\theta_N})$.
The feasible IRS configuration set is $\cQ\triangleq\Phi_K^N$.
Denote by $\bm G\in\mathbb C^{N\times N_t}$ the channel from BS to the IRS and
denote by $\bm h_{r,u}\in\mathbb C^N$ the channel from the IRS to user $u$. We assume that the direct links between BS and users are blocked. Then, the channel from the BS to the IRS and then to user $u$ is given by
\begin{equation}
 \bm h_u^H(\bm\theta)=\bm h_{r,u}^H\bm\Theta(\bm\theta)\bm G,
  \label{eq:effective_channel}
\end{equation}
and we denote by
\begin{equation}
\bm H(\bm\theta)=[\bm h_1(\bm\theta),\ldots,
 \bm h_U(\bm\theta)]^H.
\end{equation}
the corresponding channel matrix.
We adopt a block-fading model, with the channel constant within each coherence block and independent across blocks.

Before data transmission, the legitimate controller obtains
$\bm\theta^{(0)}$ using an existing IRS configuration method.
Let $\bm H_0=\bm H(\bm\theta^{(0)})$. Assuming
$N_t\geq U$ and $\operatorname{rank}(\bm H_0)=U$, the BS constructs the
power-normalized ZF precoder $\bm W$ \cite{heath2018mimo}.
Denote by $\bm s=[s_1,s_2,\ldots,s_U]^\mathrm{T}$ the vector of user-specific symbols, where each individual symbol $s_u$ is drawn from an $M$-PSK constellation:
\begin{equation}
\label{eq:mpsk_constellation}
s_u\in\mathcal S_M
=\left\{e^{j\frac{2\pi m}{M}} \mid m=0,1,\ldots,M-1\right\}.
\end{equation}
In the absence of the attacker, the configuration of IRS retains
$\bm\theta^{(0)}$, the BS transmits $\bm x=\bm W\bm s$, and user $u$
receives
\begin{equation}
 y_u^{(0)}=\bm h_u^H(\bm\theta^{(0)})\bm W\bm s+n_u
 =\rho_u s_u+n_u,\;\rho_u>0,
 \label{eq:legitimate_received}
\end{equation}
where $n_u\sim\CN(0,\sigma^2)$ is the Gaussian noise and the second equality follows from ZF suppression of interuser interference. We adopt a transmitter-side CSI architecture: CSI is used by the legitimate controller and the BS to configure the IRS and construct the ZF precoder, whereas the user terminals contain no effective-channel estimation or channel-equalization stage. Under the legitimate IRS configuration, transmitter-side ZF yields the positive real-valued gain $\rho_u$, so each user directly performs symbol-by-symbol nearest-neighbor detection:
\begin{equation}
 \widehat s_u=\arg\min_{s\in\mathcal S_M}|y_u^{(0)}-s|^2.
 \label{eq:detector}
\end{equation}

We now describe the attack model. After the legitimate IRS configuration $\bm\theta^{(0)}$ and ZF precoder $\bm W$ have been established, an attacker controls the IRS and tests $\bm\theta\neq\bm\theta^{(0)}$ during payload transmission. All nodes and surrounding scatterers remain stationary during each attack episode, so the complete search, including BER observation and feedback, fits within one coherence block. The tests may span multiple payload frames, but the users continue to perform the nearest-neighbor detection in \eqref{eq:detector}. Meanwhile, because the legitimate system does not detect the IRS reconfiguration, the BS does not restore or redesign $\bm\theta^{(0)}$ and does not update $\bm W$. The fixed precoder therefore no longer zero-forces $\bm H(\bm\theta)$, leaving uncompensated desired-signal distortion and residual interuser interference.

We now formulate the attack model mathematically. For a fixed channel and phase vector $\bm\theta$, let
$p_u(\bm\theta)$ denote the true BER of user $u$. In practice, each user
computes and reports the corresponding BER estimate
$\widehat p_u(\bm\theta)$ according to ITU-T Recommendation O.151
\cite{itut1992o151}. We assume that the attacker has implanted a backdoor in
the user devices through a malicious broadcast message, allowing it to
obtain their BER reports without being detected. At query $t$, the attacker
applies $\bm\theta^{(t)}$ and observes the reported BER vector. We define the
true objective, the observation vector, and the observed score as
\begin{equation}
\begin{aligned}
 f(\bm\theta)
 &\triangleq \min_{1\leq u\leq U}p_u(\bm\theta),\\
 \widehat{\bm p}^{(t)}
 &\triangleq [\widehat p_u(\bm\theta^{(t)})]_{u=1}^{U},\\
 \widehat f(\bm\theta^{(t)})
 &\triangleq \min_{1\leq u\leq U}\widehat p_u(\bm\theta^{(t)}).
\end{aligned}
\label{eq:utility}
\end{equation}
Given a query budget $B\leq K^N$, the attacker sequentially tests $B$
distinct phase vectors in $\Phi_K^N$, where each query is selected using
only the previously reported BER vectors. After the $B$ queries, the
attacker returns one of the tested configurations, denoted by
$\bm\theta^\star$. The goal is to maximize the expected true minimum BER
$\E[f(\bm\theta^\star)]$, although only the noisy scores
$\widehat f(\bm\theta^{(t)})$ are observable during the search.

\section{Proposed Progressive Reward-Informed Search Method}
\label{sec:method}

PRIS performs a history-guided coarse-to-fine search using only BER
feedback. It maintains a current IRS configuration, applies a sampled block
of phase changes to form a complete candidate, and physically queries that
candidate once. Large blocks explore distant configurations early in the
search, while single-element changes provide local refinement later. The
returned BER vector updates the proposal policy, and an acceptance rule
selects the starting configuration for the next query.

The initial configuration is sampled uniformly from $\cQ$. At query $t$,
PRIS changes $b_t$ distinct elements of $\bm\theta_t^{\rm cur}$ through
$\bm a_t=\{(n_{t,j},\ell_{t,j})\}_{j=1}^{b_t}$ and obtains
$\bm\theta_t^{\rm cand}$. Candidate generation is guided by two zero-initialized
preference tables:
\begin{equation}
\bm A\in\mathbb R^{N\times K},\;
\bm C\in\mathbb R^{N\times K\times K}.
\label{eq:tables}
\end{equation}
Here, $A_{n,k}$ scores selecting element $n$ at phase $k$, whereas
$C_{n,k,\ell}$ scores changing it from $k$ to $\ell$. With
$\mathcal U_{t,j}$ denoting the unselected elements, PRIS samples
\begin{equation}
 \pi^{\rm sel}_{t,j}(n)=(1-\xi_t)
 \frac{e^{A_{n,\theta_{t,n}^{\rm cur}}/\tau_t}}
 {\sum_{i\in\mathcal U_{t,j}}e^{A_{i,\theta_{t,i}^{\rm cur}}/\tau_t}}
 +\frac{\xi_t}{|\mathcal U_{t,j}|}.
 \label{eq:element_policy}
\end{equation}
After selecting $n$, its new phase is sampled from
\begin{equation}
 \pi^{\rm ph}_{t,j}(\ell)=(1-\xi_t)
 \frac{e^{C_{n,\theta_{t,n}^{\rm cur},\ell}/\tau_t}}
 {\sum_{c\in\mathcal L_{t,n}}e^{C_{n,\theta_{t,n}^{\rm cur},c}/\tau_t}}
 +\frac{\xi_t}{K-1}.
 \label{eq:phase_policy}
\end{equation}
where $\mathcal L_{t,n}=\Phi_K\setminus\{\theta_{t,n}^{\rm cur}\}$.
The action probability $\pi_t$ is the product of these selection and phase
probabilities. The exploration mixture $\xi_t$ and policy temperature
$\tau_t$ are linearly annealed within each stage.

Define normalized progress $\zeta_t=(t-2)/\max(B-2,1)$. The main
coarse-to-fine schedule is
\begin{equation}
b_t=\begin{cases}
\max\{2,\lfloor\bar b(\zeta_t)+\tfrac12\rfloor\},&\zeta_t<\gamma,\\
1,&\zeta_t\geq\gamma.
\end{cases}
\label{eq:block_schedule}
\end{equation}
where $\bar b(\zeta)$ decreases linearly over the three intervals ending at
$\rho_1$, $\rho_2$, and $\gamma$. Both stages share the same tables.

To balance users during block exploration, PRIS learns from an annealed soft
minimum. For $p_{\min}=\min_u p_u$,
\begin{equation}
 g_{\zeta}(\bm p)=
 \left\{
 \begin{array}{@{}l@{\;}l@{}}
 p_{\min}-\lambda_{\zeta}\log\!\left(\dfrac{1}{U}
 \displaystyle\sum_{u=1}^{U}e^{-(p_u-p_{\min})/\lambda_{\zeta}}\right),
 &\zeta<\gamma,\\[1ex]
 p_{\min},&\zeta\geq\gamma,
 \end{array}
 \right.
 \label{eq:softmin}
\end{equation}
The temperature $\lambda_\zeta$ decreases until refinement, where
$g_\zeta$ becomes the hard minimum. Let $\Delta_t^{\rm learn}$ be the
change in $g_{\zeta_t}$ and $\Delta_t^{\rm obj}$ the change in the observed
minimum BER. With $\bm\vartheta=\operatorname{vec}(\bm A,\bm C)$, the policy
is updated by Adam ascent using the normalized direction
$w_t\nabla_{\bm\vartheta}\log\pi_t(\bm a_t\mid\bm\theta_t^{\rm cur})$,
where $w_t$ is a bounded, noise-normalized version of
$\Delta_t^{\rm learn}$. Every $I_{\rm r}$ queries, at most $R_{\max}$
successful transitions are replayed under the current policy without new
physical queries. Because no importance correction is applied, replay is a
deliberately biased reinforcement heuristic.

The objective improvement controls only the next search state; every tested
candidate remains eligible for final selection. The acceptance probability is
\begin{equation}
P_{{\rm acc},t}=
\begin{cases}
1,&\Delta_t^{\rm obj}\geq0,\\
p_{\rm down}(\zeta_t),
&\Delta_t^{\rm obj}<0,\ \zeta_t<\gamma,\\
\exp\!\left[\dfrac{\Delta_t^{\rm obj}}{T_{\rm acc}(\zeta_t)}\right],
&\Delta_t^{\rm obj}<0,\ \zeta_t\geq\gamma.
\end{cases}
\label{eq:acceptance}
\end{equation}
Here, $p_{\rm down}$ decreases during exploration, while
$T_{\rm acc}$ is exponentially annealed from $T_0$ to $T_f$.
Finally, a hash set prevents repeated tests: duplicate proposals are resampled
by rejection from $\cQ\setminus\cV_{t-1}$. PRIS returns
$\bm\theta^{\mathrm{PRIS}}\in\arg\max_{1\leq t\leq B}
\widehat f(\bm\theta^{(t)})$. Its controller time and storage complexities
are $O(B[NK^2+N(N+K)])$ and $O(NK^2+BN+BU)$, respectively.

\begin{figure*}[!t]
\centering
\compactfigurecaptions

\begin{minipage}[t]{0.48\textwidth}
\centering
\includegraphics[width=0.88\linewidth,trim=0bp 0bp 0bp 20bp,clip]{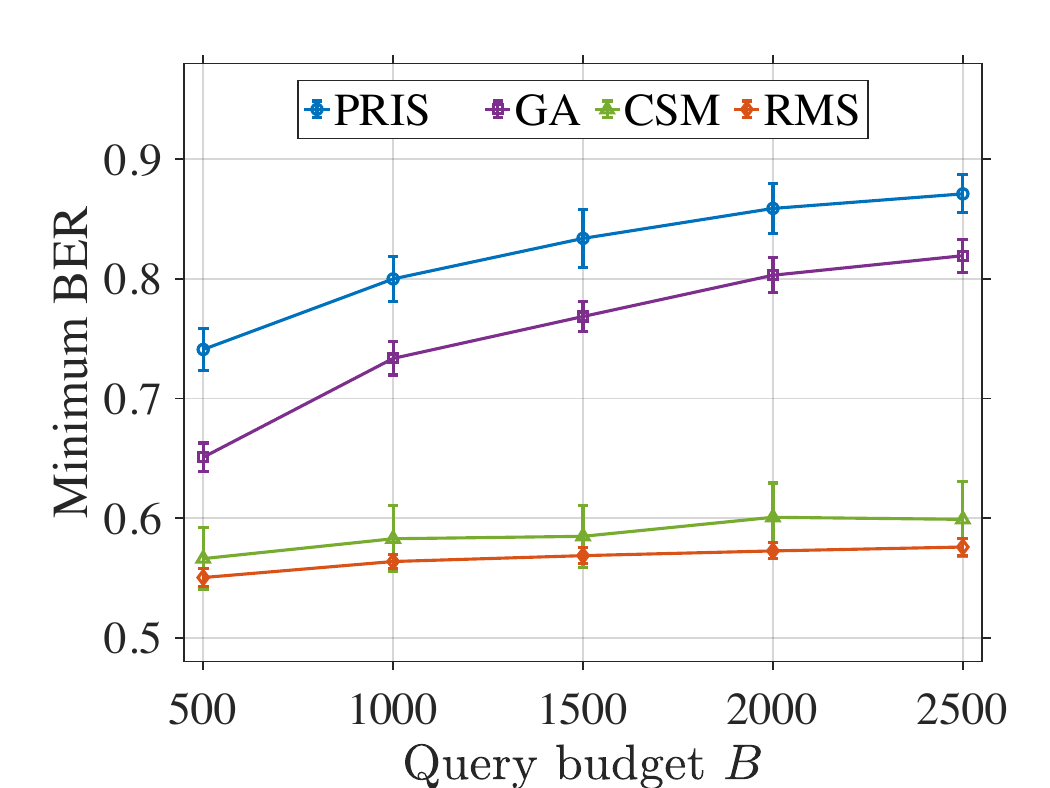}
\captionof{figure}{BER versus query budget.}
\label{fig:budget_results}
\end{minipage}\hfill
\begin{minipage}[t]{0.48\textwidth}
\centering
\includegraphics[width=0.88\linewidth,trim=0bp 0bp 0bp 20bp,clip]{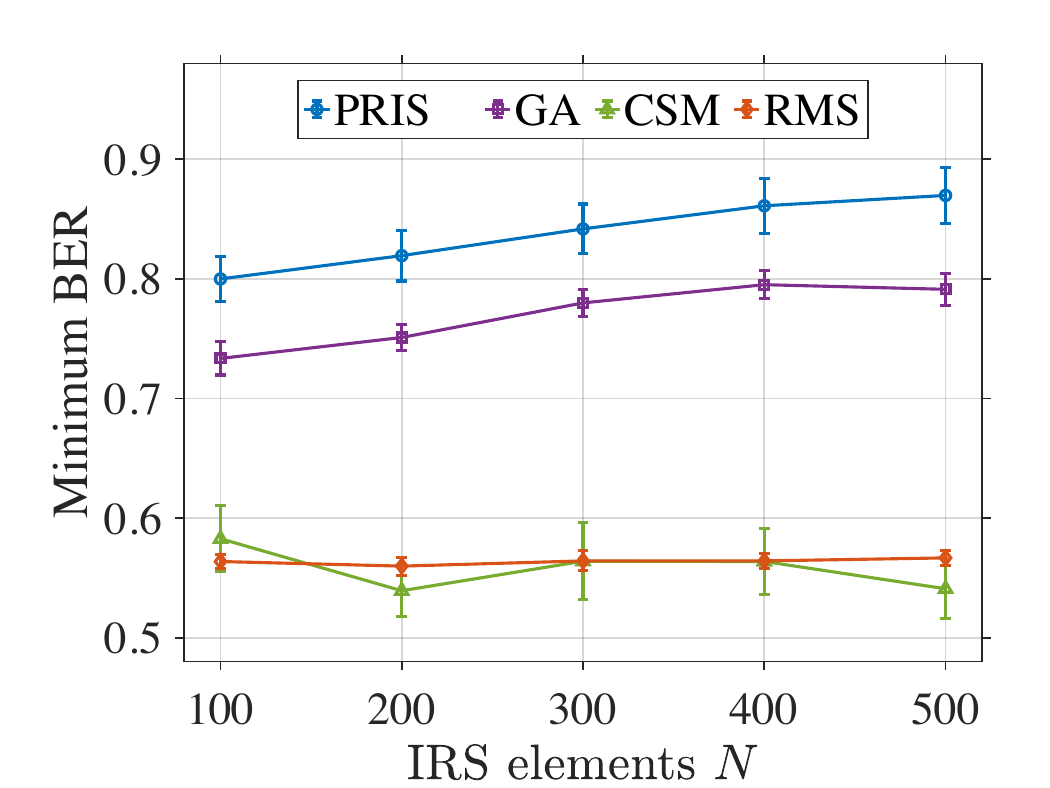}
\captionof{figure}{BER versus IRS size ($B=10N$).}
\label{fig:size_results}
\end{minipage}

\par\vspace{2pt}

\begin{minipage}[t]{0.48\textwidth}
\centering
\includegraphics[width=0.88\linewidth,trim=0bp 0bp 0bp 20bp,clip]{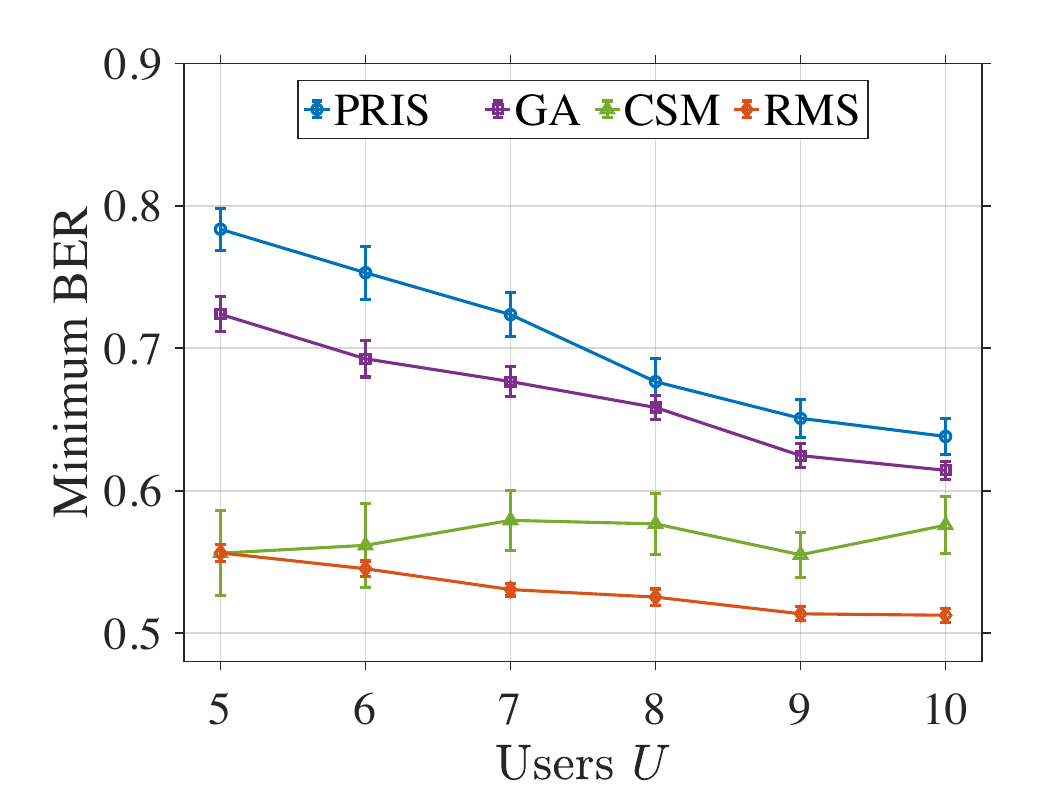}
\captionof{figure}{BER versus user count.}
\label{fig:user_results}
\end{minipage}\hfill
\begin{minipage}[t]{0.48\textwidth}
\centering
\includegraphics[width=0.88\linewidth,trim=0bp 0bp 0bp 20bp,clip]{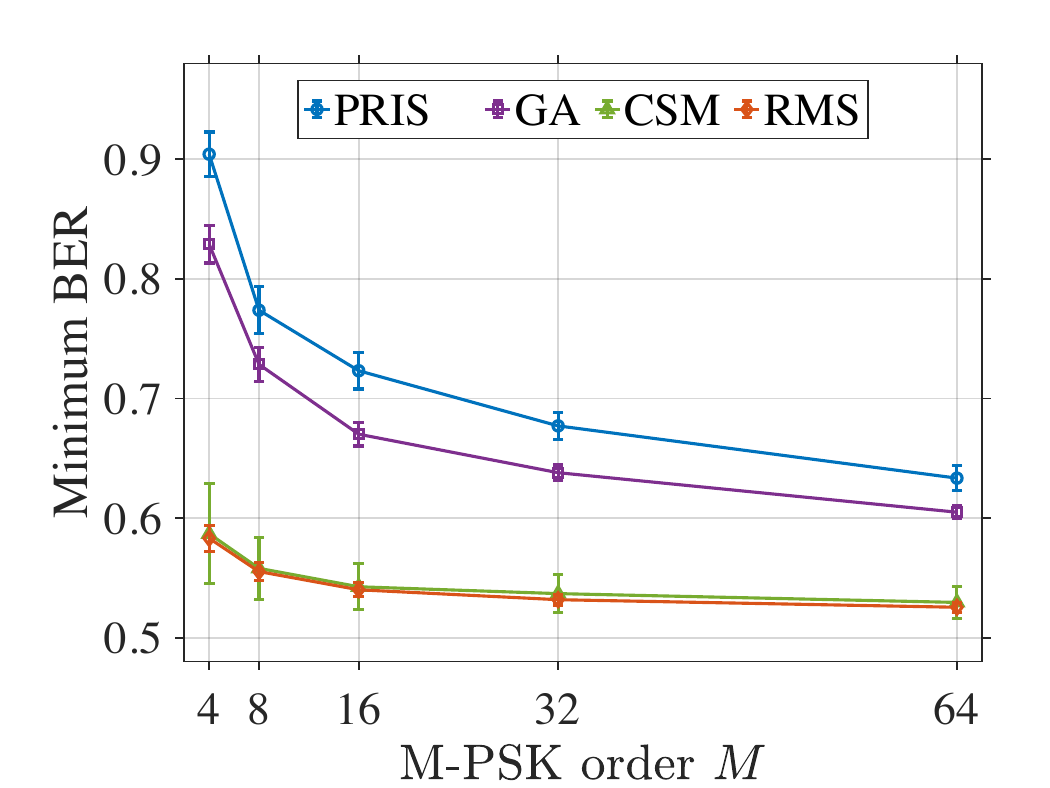}
\captionof{figure}{BER versus M-PSK order.}
\label{fig:modulation_results}
\end{minipage}
\end{figure*}

\section{Experiments}
\label{sec:experiments}

\subsection{Simulation Setup}
We evaluate the proposed PRIS through numerical simulations. Unless otherwise stated, the BS has $N_t=32$ antennas and is located at $(50,-50,10)$~m, while the IRS is located at $(-2,-1,0)$~m. Each RE of the IRS has four discrete phase shift options: $\{0, \pi/2, \pi, 3\pi/2\}$. The users are independently placed in the $5\times5$~m$^2$ region next to the IRS. The default numbers of RE and users are $N=100$ and $U=5$, respectively. The BS--IRS and IRS--user channels follow geometry-dependent Rician fading at 3.5~GHz with Rician factors of 10 and 3~dB, respectively. Both links use a path-loss exponent of 2.2 and a reference loss of 30~dB at 1~m. The BS transmit power and receiver noise power are 30 and $-90$~dBm, respectively. We use Gray-mapped 8-PSK for data transmission. Before data transmission, we assume that the legitimate controller adopts two sweeps of discrete coordinate ascent in \cite{wu2020discrete} to configure the IRS.

We compare the proposed PRIS with three benchmarks:
\begin{itemize}
    \item \emph{Genetic Algorithm (GA) \cite{goldberg1989genetic}:} It performs the conventional GA algorithm with a population of 50.
    \item \emph{Conditional Sample Mean (CSM) \cite{ren2023blind}:} It tries out a sequence of random IRS configurations and optimizes the configuration to maximize the conditional sample mean of BER.
    \item \emph{Random Max Sampling (RMS) \cite{ren2023blind}:} It tries out a sequence of random IRS configurations and chooses the best that maximizes the finite-frame BER.
\end{itemize}

For the proposed PRIS, we set the parameters $(\rho_1,\rho_2,\gamma)\allowbreak=\allowbreak(0.30,0.55,0.75)$. Across the three
intervals, the block size decreases from $0.30N$ to $0.20N$, from
$0.15N$ to $0.08N$, and from $0.04N$ to $0.02N$. Adam uses $\eta=0.01$;
replay is performed every $I_{\rm r}=2$ queries with $R_{\max}=16$;
and refinement uses $(T_0,T_f)=(0.030,0.002)$. After performing the proposed PRIS or the benchmarks, the selected configuration is evaluated on an independent frame of $100000$ symbols to approximate the true BER.

\subsection{Experiment Results}

We evaluate the performance of different algorithms under different network parameter settings. Among these parameters, the query budget has the most significant impact on performance.
Fig.~\ref{fig:budget_results} shows the minimum BER versus the number of queries $B$. It can be seen that the minimum BER achieved by all methods increases with $B$. As $B$ increases from 500 to 2500, the minimum BER of PRIS rises from $74.1\%$ to $87.1\%$. Compared to GA, the advantage of PRIS is $9\%$ at $B=500$ and $5.2\%$ at $B=2500$. In contrast, CSM and RMS improve only slightly when $B$ increases. This indicates that the query efficiency of these methods is low, as even a large query budget fails to yield satisfactory performance.

Fig.~\ref{fig:size_results} further shows the performance of different algorithms under different numbers of REs $N$ when $B=10N$. Observe that the minimum BER achieved by PRIS improves from $0.8$ at $N=100$ to $0.87$ at $N=500$ and maintains a margin of $0.062$--$0.078$ over GA. In contrast, the performance of CSM and RMS does not change with $N$. Fig.~\ref{fig:user_results} then shows the minimum BER versus the number of users $U$ when $N=100$ and $B=1000$. We find that the minimum BER of almost all methods decreases as it becomes more challenging to cause decoding errors for a larger number of users. Nevertheless, PRIS consistently achieves the best performance regardless of the number of users.

We next vary the M-PSK order over $M\in\{4,8,16,\allowbreak32,64\}$ to evaluate the performance of different algorithms. As shown in Fig.~\ref{fig:modulation_results}, the minimum BER decreases for every method as the constellation order increases. PRIS nevertheless remains the best method at every tested order. Its mean advantage over GA is $0.075$ for QPSK and remains $0.028$ for 64-PSK, while its performance advantage over CSM and RMS remains substantially larger.

\section{Conclusion}
\label{sec:conclusion}
This paper investigated a BER-only black-box attack on an IRS-assisted
multiuser MISO downlink, where an attacker reconfigures the IRS to degrade
legitimate data transmission. With access to neither channel information
nor symbol-level observations, the attacker must identify an IRS
configuration that maximizes the minimum BER across users using only
finite-frame BER feedback and a limited budget of non-repeated tests.
To address this problem, we propose PRIS, which learns effective phase
transitions from previous trials. Its search progressively moves from
large-block exploration to single-element refinement. An annealed
user-balancing reward and prioritized replay are incorporated to exploit
noisy BER feedback. The numerical evaluation considers different query
budgets, IRS sizes, user counts, and M-PSK modulation orders. Comparisons
with GA, CSM, and RMS demonstrate that PRIS achieves a higher minimum BER
under the same query budget and remains effective across the investigated
network settings.

\clearpage
\begingroup
\fontsize{10}{10.7}\selectfont
\let\originalthebibliography\thebibliography
\renewcommand{\thebibliography}[1]{%
  \originalthebibliography{#1}%
  \setlength{\itemsep}{0.5pt}%
  \setlength{\parsep}{0pt}%
  \setlength{\parskip}{0pt}%
}
\bibliographystyle{IEEEbib}
\bibliography{IEEEabrv, references}

@article{basar2019ris,
  author  = {E. Basar and M. Di Renzo and J. de Rosny and M. Debbah and M.-S. Alouini and R. Zhang},
  title   = {Wireless communications through reconfigurable intelligent surfaces},
  journal = {IEEE Access},
  volume  = {7},
  pages   = {116753--116773},
  year    = {2019}
}

@article{direnzo2020smart,
  author  = {M. Di Renzo and A. Zappone and M. Debbah and M.-S. Alouini and C. Yuen and J. de Rosny and S. Tretyakov},
  title   = {Smart radio environments empowered by reconfigurable intelligent surfaces: How it works, state of research, and the road ahead},
  journal = {IEEE J. Sel. Areas Commun.},
  volume  = {38},
  number  = {11},
  pages   = {2450--2525},
  month   = nov,
  year    = {2020}
}

@article{wu2021tutorial,
  author  = {Q. Wu and S. Zhang and B. Zheng and C. You and R. Zhang},
  title   = {Intelligent reflecting surface-aided wireless communications: A tutorial},
  journal = {IEEE Trans. Commun.},
  volume  = {69},
  number  = {5},
  pages   = {3313--3351},
  month   = may,
  year    = {2021}
}

@article{wu2019joint,
  author  = {Q. Wu and R. Zhang},
  title   = {Intelligent reflecting surface enhanced wireless network via joint active and passive beamforming},
  journal = {IEEE Trans. Wireless Commun.},
  volume  = {18},
  number  = {11},
  pages   = {5394--5409},
  month   = nov,
  year    = {2019}
}

@article{guo2020wsr,
  author  = {H. Guo and Y.-C. Liang and J. Chen and E. G. Larsson},
  title   = {Weighted sum-rate maximization for reconfigurable intelligent surface aided wireless networks},
  journal = {IEEE Trans. Wireless Commun.},
  volume  = {19},
  number  = {5},
  pages   = {3064--3076},
  month   = may,
  year    = {2020}
}

@article{pan2020multicell,
  author  = {C. Pan and H. Ren and K. Wang and W. Xu and M. Elkashlan and A. Nallanathan and L. Hanzo},
  title   = {Multicell {MIMO} communications relying on intelligent reflecting surfaces},
  journal = {IEEE Trans. Wireless Commun.},
  volume  = {19},
  number  = {8},
  pages   = {5218--5233},
  month   = aug,
  year    = {2020}
}

@article{wu2020discrete,
  author  = {Q. Wu and R. Zhang},
  title   = {Beamforming optimization for wireless network aided by intelligent reflecting surface with discrete phase shifts},
  journal = {IEEE Trans. Commun.},
  volume  = {68},
  number  = {3},
  pages   = {1838--1851},
  month   = mar,
  year    = {2020}
}

@article{ren2023blind,
  author  = {S. Ren and K. Shen and Y. Zhang and X. Li and X. Chen and Z.-Q. Luo},
  title   = {Configuring intelligent reflecting surface with performance guarantees: Blind beamforming},
  journal = {IEEE Trans. Wireless Commun.},
  volume  = {22},
  number  = {5},
  pages   = {3355--3370},
  month   = may,
  year    = {2023}
}

@article{lyu2020jamming,
  author  = {B. Lyu and D. T. Hoang and S. Gong and D. Niyato and D. I. Kim},
  title   = {{IRS}-based wireless jamming attacks: When jammers can attack without power},
  journal = {IEEE Wireless Commun. Lett.},
  volume  = {9},
  number  = {10},
  pages   = {1663--1667},
  month   = oct,
  year    = {2020}
}

@article{wang2022illegal,
  author  = {Y. Wang and H. Lu and D. Zhao and Y. Deng and A. Nallanathan},
  title   = {Wireless communication in the presence of illegal reconfigurable intelligent surface: Signal leakage and interference attack},
  journal = {IEEE Wireless Commun.},
  volume  = {29},
  number  = {3},
  pages   = {131--138},
  month   = jun,
  year    = {2022}
}

@article{huang2021pilot,
  author  = {K.-W. Huang and H.-M. Wang},
  title   = {Intelligent reflecting surface aided pilot contamination attack and its countermeasure},
  journal = {IEEE Trans. Wireless Commun.},
  volume  = {20},
  number  = {1},
  pages   = {345--359},
  month   = jan,
  year    = {2021}
}

@article{yang2021spoofing,
  author  = {J. Yang and X. Ji and F. Wang and K. Huang and L. Guo},
  title   = {A novel pilot spoofing scheme via intelligent reflecting surface based on statistical {CSI}},
  journal = {IEEE Trans. Veh. Technol.},
  volume  = {70},
  number  = {12},
  pages   = {12847--12857},
  month   = dec,
  year    = {2021}
}

@inproceedings{li2025gadrl,
  author    = {Z. Li and Y. Ju and J. Hu and L. Liu and Y. G. Shee and C. Wu and S. Mumtaz},
  title     = {Wireless communication attacks based on {GA--DRL} assisted non-diagonal {RIS}},
  booktitle = {Proc. IEEE INFOCOM Workshops},
  pages     = {1--6},
  year      = {2025},
  doi       = {10.1109/INFOCOMWKSHPS65812.2025.11152859}
}

@inproceedings{mughal2025malris,
  author    = {D. M. Mughal and D. Munir and Q. A. Ahmed and H. D. Schotten and T. Jungeblut and S.-H. Kim and M. Y. Chung},
  title     = {{MALRIS}: Malicious hardware in {RIS}-assisted wireless communications},
  booktitle = {Proc. IEEE CSCN},
  year      = {2025},
  doi       = {10.1109/CSCN67557.2025.11230594}
}

@book{heath2018mimo,
  author    = {R. W. Heath Jr. and A. Lozano},
  title     = {Foundations of {MIMO} Communication},
  publisher = {Cambridge University Press},
  address   = {Cambridge, U.K.},
  year      = {2018}
}

@book{goldberg1989genetic,
  author    = {D. E. Goldberg},
  title     = {Genetic Algorithms in Search, Optimization, and Machine Learning},
  publisher = {Addison-Wesley},
  address   = {Reading, MA, USA},
  year      = {1989}
}

@techreport{itut1992o151,
  author      = {{ITU-T}},
  title       = {Error Performance Measuring Equipment Operating at the Primary Rate and Above},
  institution = {International Telecommunication Union},
  number      = {Recommendation O.151},
  month       = oct,
  year        = {1992}
}

@article{dinhvan2025born,
  author  = {S. Dinh-Van and N. P. Tran and M. D. Higgins},
  title   = {Near-optimal reconfigurable intelligent surface configuration: Blind beamforming with sensing},
  journal = {arXiv preprint arXiv:2511.05132},
  year    = {2025}
}

@article{lai2025blindmimo,
  author  = {W. Lai and J. Yao and K. Shen},
  title   = {Blind passive beamforming for {MIMO} system},
  journal = {IEEE Wireless Commun. Lett.},
  volume  = {14},
  number  = {8},
  pages   = {2601--2605},
  month   = aug,
  year    = {2025},
  doi     = {10.1109/LWC.2025.3576288}
}

@article{gong2020toward,
  title={Toward smart wireless communications via intelligent reflecting surfaces: A contemporary survey},
  author={S. Gong and X. Lu and D. T. Hoang and D. Niyato and L. Shu and D. I. Kim and Y.-C. Liang},
  journal={IEEE Commun. Surveys Tuts.},
  volume={22},
  number={4},
  pages={2283--2314},
  year={2020}
}

@article{huang2019reconfigurable,
  title="Reconfigurable intelligent surfaces for energy efficiency in wireless communication",
  author={C. Huang and A. Zappone and G. C. Alexandropoulos and M. Debbah and C. Yuen},
  journal={IEEE Trans. Wireless Commun.},
  volume=18,
  number=8,
  pages={4157--4170},
  year=2019,
  month=aug
}

@article{liu2021reconfigurable,
  title={Reconfigurable intelligent surfaces: Principles and opportunities},
  author={Y. Liu and X. Liu and X. Mu and T. Hou and J. Xu and M. Di Renzo and N. Al-Dhahir},
  journal={IEEE Commun. Surveys Tuts.},
  volume={23},
  number={3},
  pages={1546--1577},
  year={2021}
}

@article{huang2026discoios,
  author  = {H. Huang and H. Zhang and J. Yuan and L. Sun and Y. Wang and W. Mei and B. Di and Y. Cai and Z. Han},
  title   = {Disco intelligent omni-surfaces: {$360^\circ$} fully-passive jamming attacks},
  journal = {IEEE Trans. Wireless Commun.},
  volume  = {25},
  pages   = {61--74},
  year    = {2026},
  doi     = {10.1109/TWC.2025.3581208}
}

@article{shao2023secure,
  author  = {R. Shao and T. Li and Y. Song and W. Ji and F. Li},
  title   = {A secure physical layer transport scheme with illegal reconfigurable intelligent surface},
  journal = {AEU Int. J. Electron. Commun.},
  volume  = {170},
  note    = {Art. no. 154837},
  month   = oct,
  year    = {2023},
  doi     = {10.1016/j.aeue.2023.154837}
}

@article{yang2021antijamming,
  author  = {H. Yang and Z. Xiong and J. Zhao and D. Niyato and Q. Wu and H. V. Poor and M. Tornatore},
  title   = {Intelligent reflecting surface assisted anti-jamming communications: A fast reinforcement learning approach},
  journal = {IEEE Trans. Wireless Commun.},
  volume  = {20},
  number  = {3},
  pages   = {1963--1974},
  month   = mar,
  year    = {2021},
  doi     = {10.1109/TWC.2020.3037767}
}

@article{sun2022outage,
  author  = {Y. Sun and K. An and J. Luo and Y. Zhu and G. Zheng and S. Chatzinotas},
  title   = {Outage constrained robust beamforming optimization for multiuser {IRS}-assisted anti-jamming communications with incomplete information},
  journal = {IEEE Internet Things J.},
  volume  = {9},
  number  = {15},
  pages   = {13298--13314},
  month   = aug,
  year    = {2022},
  doi     = {10.1109/JIOT.2022.3140752}
}

@article{wang2026distributeddisco,
  author  = {Y. Wang and S. Li and H. Huang and Y. Zhang and L. Sun and Y. Song and J. Yuan and T. Yu and Y. Cai},
  title   = {Distributed disco intelligent reflecting surfaces-based fully passive jamming for {MU-MISO} systems},
  journal = {Electronics},
  volume  = {15},
  number  = {10},
  note    = {Art. no. 2033},
  year    = {2026},
  doi     = {10.3390/electronics15102033}
}
\endgroup

\end{document}